\documentstyle[12pt]{article}

\newcommand{\mysection}{\setcounter{equation}{0}\section}

\begin{document}

\hfill{ITP-SB-93-46} 
\vskip 0.1cm
\hfill{SMU HEP 93-14} 
\vskip 0.1cm
\hfill{FERMILAB-Pub-93/240-T} 
\vskip 0.8cm
\centerline{\large\bf { Complete Next to Leading Order QCD Corrections}}
\vskip 0.2cm 
\centerline{\large\bf{to the Photon Structure Functions
 $F^\gamma_2(x,Q^2)$ and $F_L^\gamma(x,Q^2)$}}
\vskip 0.7cm
\centerline{\sc E. Laenen,}
\vskip 0.3cm
\centerline{\it Fermi National Accelerator Laboratory,} 
\centerline{\it P.O. Box 500, MS 106}
\centerline{\it Batavia, Illinois 60510}
\vskip 0.3cm
\centerline {\sc S. Riemersma, }
\vskip 0.3cm
\centerline{\it Department of Physics,}
\centerline{\it Fondren Science Building,}
\centerline{\it Southern Methodist University,}
\centerline{\it Dallas, Texas 75275}
\vskip 0.3cm 
\centerline {\sc J. Smith, }
\vskip 0.3cm
\centerline{\it Institute for Theoretical Physics,}
\centerline{\it State University of New York at Stony Brook,}
\centerline{\it Stony Brook, New York 11794-3840}
\vskip 0.3cm 
\centerline{and}
\vskip 0.3cm 
\centerline{\sc W. L. van Neerven}
\vskip 0.3cm 
\centerline{\it Instituut Lorentz,}
\centerline{\it University of Leiden,}
\centerline{\it P.O.B. 9506, 2300 RA, Leiden,} 
\centerline{\it The Netherlands.}
\vskip 0.3cm
\centerline{August 1993}
\vskip 0.3cm
\newpage
\centerline{\bf Abstract}
\vskip 0.4cm

We present the complete NLO QCD analysis of the photon
structure functions $F_2^\gamma(x,Q^2)$ and $F_L^\gamma(x,Q^2)$
for a real photon target.
In particular we study the heavy flavor content of the
structure functions which is due to two different 
production mechanisms, namely collisions of a virtual photon 
with a real photon, and with a  parton. 
We observe that the charm contributions are noticeable
for $F_2^\gamma(x,Q^2)$ as well as
$F_L^\gamma(x,Q^2)$ in the x-region studied.

\vfill
\newpage
\mysection{Introduction}
In the past two decades there has been considerable interest
in the study of photon-photon interactions in electron-positron colliders.
When one photon is virtual and the other one is almost real
the analogy with deep-inelastic electron-nucleon scattering motivates
the introduction of the corresponding structure functions 
$F_k^\gamma(x,Q^2)$ $(k=2,L)$ for the photon.  
The deep-inelastic structure function $F_2^\gamma(x,Q^2)$ 
was originally measured by the PLUTO collaboration \cite{pluto}
at PETRA using single-tag events in the reaction
$e^- + e^+ \rightarrow e^- + e^+ +$ hadrons. In the past several
years there has been a series of new measurements at PETRA, PEP and TRISTAN
by several groups, including CELLO \cite{cello},
TPC2$\gamma$ \cite{tpc}, TASSO \cite{tasso}, JADE \cite{jade},
AMY \cite{amy}, VENUS \cite{venus} and TOPAZ \cite{topaz}.
All these groups concentrated on the measurement of the light-quark 
contribution to $F_2^\gamma(x,Q^2)$. The heavy-quark component (mainly charm)
has been hard to extract due to problems identifying charmed particle
decays so its contribution to the data was sometimes removed 
according to a Monte Carlo estimate. 
In the near future higher-luminosity runs at TRISTAN
should yield some information on heavy-quark (mainly charm)
production and this is one reason that we study it here.
At this moment the available data for $F_2^\gamma(x,Q^2)$ 
are in the region
$0.03 < x < 0.8$ and $1.31 \, ({\rm GeV}/c)^2 <Q^2 < 390 \,
({\rm GeV}/c)^2$.
Due to the experimental limitation that $xy^2 << 1 $ (for a definition of $x$
and $y$ see (2.5)), there are no data available for the
longitudinal structure function $F^\gamma_L(x,Q^2)$. However
there exists some hope that $F_L^\gamma(x,Q^2)$ can be
measured \cite{ali} at LEP.
Finally two-photon reactions are important to understand as background
processes to the normal $s$-channel reactions at present and future 
$e^+e^-$ colliders. These machines will have a large amount of 
beamstrahlung \cite{dg1}, \cite{eghn}. Therefore a basic input is 
the parton density in a photon which will be modified if higher 
order pQCD corrections are included. 

As far as theory is concerned the first attempt to give a theoretical
description of the photon structure function in the context
of perturbative QCD was given by E. Witten in \cite{ew}.
He suggested that both the $x$ and the $Q^2$ dependence of these 
structure functions were calculable in pQCD at asymptotically large $Q^2$.
Thus from a theoretical point of view
this process should provide a much more thorough test of 
pQCD than the corresponding deep-inelastic scattering off a
nucleon target, where only the $Q^2$ evolution
of the structure functions is calculable. The original 
optimism subsided once it was realized that there were complications with
experimental confirmation of this prediction
at experimental (non-asymptotic) values of $Q^2$
\cite{bill}\,,\,\cite{gr}.  For recent reviews see \cite{jf}. 
In particular there is a contamination of the purely
pointlike pQCD contribution by the hadronic component
of the photon. This latter piece, which is most important at
small virtualities, is not calculable 
in pQCD and must be extracted from experimental data.
One of the approaches used is to describe this hadronic piece by parton
densities in the photon,
analogous to the parton densities in a hadronic target. 
For parameterizations see
\cite{dg}, \cite{acl}, \cite{grv1}, \cite{grv2}, \cite{acfgp}
and \cite{gs}. 
For a different approach see \cite{fkp}.

In \cite{grv2} a next to leading order (NLO) analysis
was carried out for the photon structure function 
$F_2^\gamma(x,Q^2)$. This analysis also includes the lowest order
contribution coming from heavy flavor production, which is described
by the Bethe-Heitler cross section corresponding to the
process $\gamma^* + \gamma \rightarrow Q + \bar Q$. In this case
the mass $m$ of the heavy flavor is not neglected with respect to
$Q^2$ especially in the threshold region. If $Q^2 >> m^2$
one encounters large logarithmic terms containing
$\ln(Q^2/m^2)$, which have to be summed
using the Altarelli-Parisi (AP) equations. This procedure 
provides us with the heavy flavor densities 
in the photon which are akin to the parton densities originating
from the light quarks in the photon.
The same procedure has been applied for the 
longitudinal structure functions $F_L^\gamma(x,Q^2)$
in \cite{dgr} but only in leading order.

In this paper we want to extend the above analysis
by including higher order pQCD corrections which were
not considered in the literature so far. Since the NLO
QCD corrections to the longitudinal
coefficient functions due to massless partons \cite{zn}
and heavy flavors \cite{lrsn1} have been recently calculated we are
now also able to present a NLO analysis for $F_L^\gamma(x,Q^2)$.
In addition we can also improve our knowledge of the heavy
flavor content of $F_2^\gamma(x,Q^2)$ by including the order $\alpha_s$
corrections to the Bethe-Heitler process 
$\gamma^* + \gamma \rightarrow Q + \bar Q$. 
We also include corrections to $F_k^\gamma (x,Q^2)$ $(k=2,L)$
due to heavy flavor production mechanisms given by the processes 
$\gamma^* + g \rightarrow Q + \bar Q$ (corrected up to 
order $\alpha_s^2$ ) and $\gamma^* + q(\bar q) \rightarrow
Q + \bar Q + q(\bar q)$, where the incoming gluon and 
(anti)quark originate from the on-mass-shell photon. Furthermore
we use the most recent gluon and (anti)quark
densities in our analysis.

Finally we should mention that there was a 
previous investigation of pQCD corrections to heavy quark
production in \cite{hr}, where it was assumed that both photons
were off-mass-shell and a small value for the 
photon virtuality was chosen for generating numerical results. Since
these authors did not therefore encounter mass singularities they
had no need to perform any mass factorization. Hence their
method was different from the one we adopt.

The paper is organized as follows. In section 2 we 
present the photonic and hadronic coefficient 
functions corrected up to next to leading order in $\alpha_s$,
which are needed for the photon structure functions 
$F_k^\gamma(x,Q^2)$ $(k=2,L)$. In section 3 we show the differences between
the leading order (LO) and the next to leading order (NLO) photon
structure functions. In particular we discuss the effect of the heavy
flavor component (mainly charm) originating from the 
hadronic as well as the pointlike photon interactions.
\newpage
%

\mysection{Higher-Order Corrections to the Photon Structure Functions}
The deep-inelastic photon structure functions denoted by
$F^\gamma_k(x,Q^2)$ $(k=2,L)$ are measured in $e^- e^+ $ collisions 
via the process (see fig.1)
\begin{equation}
e^-(p_e) + e^+ \rightarrow e^-(p_e') + e^+ + X\,,
\end{equation}
where $X$ denotes any hadronic state which is allowed by quantum-number 
conservation laws. When the outgoing electron is tagged then
the above reaction is dominated by the photon-photon collision reaction
(see fig.1)
\begin{eqnarray}
\gamma^*(q) + \gamma(k) \rightarrow X\,,
\end{eqnarray}
where one of the photons is highly virtual and the other one
is almost on-mass-shell. The process (2.1) is described by the
cross section
\begin{eqnarray}
\frac{d^2\sigma}{dxdy} &=& \int dz \, z\, f_\gamma^e (z, \frac{S}{m_e^2})
\,  \frac{2\pi\alpha^2 S}{Q^4} \nonumber \\&& 
\times \left[ \{ 1 + (1-y)^2\} 
F^\gamma_2(x,Q^2) -y^2 F^\gamma_L(x,Q^2) \right]\,,
\end{eqnarray} 
where $F^\gamma_k(x,Q^2)$ $(k=2,L)$ denote the deep-inelastic
photon structure functions and $\alpha = e^2/4\pi$
is the fine structure constant.
Furthermore the off-mass-shell photon and the on-mass-shell photon
are indicated by the four-momenta $q$ and $k$ 
respectively with $q^2 = -Q^2 <0$
and $k^2 \approx 0$. Because the photon with momentum $k$ is 
almost on-mass-shell, expression (2.3) is written in the Weizs\"acker-Williams
approximation. In this approximation the function $f^e_\gamma(z,S/m_e^2)$
is the probability of finding a photon $\gamma(k)$ in the positron,
(see fig.1). The fraction of the energy of the positron 
carried off by the photon is denoted by $z$ while
$\sqrt{S}$ is the c.m. energy of the electron-positron system.
The function $f^e_\gamma(z,S/m_e^2)$ is given by (see \cite{sn})
\begin{equation}
f^e_\gamma(z,\frac{S}{m_e^2}) = \frac{\alpha}{2\pi}
\frac{1 + (1-z)^2}{z} \ln \frac{(1-z)(zS-4m^2)}{z^2m_e^2}\,,
\end{equation}
provided a heavy quark with mass $m$ is produced.
The scaling variables $x$ and $y$ are
defined by
\begin{equation}
x = \frac{Q^2}{2k\cdot q} \,, \qquad y = \frac{k\cdot q}{k \cdot p_e}
\,, \qquad q = p_e - p_e'\,,
\end{equation}
where $p_e$, $p_e'$ are the momenta of the incoming and outgoing
electron respectively. 
Following the procedure in \cite{ggr1} the photon structure functions
in the QCD-improved parton model have the following form 
\begin{eqnarray}
&& \frac{1}{\alpha} F_k^\gamma(x,Q^2) = 
x \int_x^1 \frac{dz}{z} \, \left[ \left(\frac{1}{n_f}
\sum_{i=1}^{n_f} e_i^2 \right)\{\Sigma^\gamma(\frac{x}{z},M^2)  
\,{\cal C}_{k,q}^S(z,\frac{Q^2}{M^2}) \right. \nonumber \\ && 
\qquad \left. +  g^\gamma(\frac{x}{z} ,M^2)
\,{\cal C}_{k,g}(z,\frac{Q^2}{M^2})\} 
+\Delta^\gamma(\frac{x}{z},M^2) \,{\cal C}_{k,q}^{NS}(z,\frac{Q^2}{M^2})
\right]\nonumber \\ && 
+ x \int_x^{z_{\rm max}} \frac{dz}{z} \, \left[ \left(\frac{1}{n_f}
\sum_{i=1}^{n_f} e_i^2 \right)\{\Sigma^\gamma(\frac{x}{z},M^2)  
\,{\cal C}_{k,q}^S(z,\frac{Q^2}{M^2},m^2) \right. \nonumber \\ && 
\qquad \left. +  g^\gamma(\frac{x}{z} ,M^2)
\,{\cal C}_{k,g}(z,\frac{Q^2}{M^2},m^2)\} 
+\Delta^\gamma(\frac{x}{z},M^2) \,{\cal C}_{k,q}^{NS}(z,
\frac{Q^2}{M^2},m^2)
\right]\nonumber \\ && 
+ x \int_x^{z_{\rm max}} \frac{dz}{z} \, \left[ 
e_H^2 \{\Sigma^\gamma(\frac{x}{z},M^2)  
\,{\cal C}_{k,q}^H(z,\frac{Q^2}{M^2},m^2) \right. \nonumber \\ && 
\qquad \left. +  g^\gamma(\frac{x}{z} ,M^2)
\,{\cal C}_{k,g}^H(z,\frac{Q^2}{M^2},m^2)\} 
\right]\nonumber \\ && 
\qquad +\frac{3}{4\pi} x \left[\left(\sum_{i=1}^{n_f} e^4_i\right) 
\,{\cal C}_{k, \gamma}(x,\frac{Q^2}{M^2})   
+ e_H^4\,{\cal C}_{k,\gamma}^H (x,Q^2,m^2)\right]
 \,.  
\end{eqnarray}
where the meaning of the symbols is explained below.

The quantities $\Sigma^\gamma$ and $\Delta^\gamma$ represent the singlet
and non-singlet combinations of the quark densities in the photon
respectively while the gluon density is represented by $g^\gamma$.
The same flavor decomposition is
also applied to the hadronic (Wilson) coefficient
functions ${\cal C}_{k,i}$ ($i=q,g$) so that 
${\cal C}_{k,q}^S(z,Q^2/M^2)$ and  ${\cal C}_{k,q}^{NS}(z,Q^2/M^2)$ stand
for the singlet and non-singlet coefficient functions respectively,
and ${\cal C}_{k,g}(z,Q^2/M^2)$ denotes the gluonic coefficient function,
where $M^2$ is the mass factorization scale.
The hadronic coefficient functions can be attributed to hard processes
with a light quark or gluon in the initial state, such as
$\gamma^* + q \rightarrow q + g$ or $\gamma^* + g \rightarrow q +
\bar q$, where the initial parton 
emerges from the real (on-mass-shell) photon. Hence they are multiplied
by the corresponding parton densities in the photon. 

We also make a distinction between light and heavy flavor
contributions to the coefficient functions. The latter are 
indicated by their explicit dependence on the heavy flavor
mass $m$. For example in the contribution to
${\cal C}_{k,i}(z,Q^2/M^2,m^2)$ (second part of (2.6)) the virtual
photon is attached either to the incoming light quark
as is the case in the reaction 
$\gamma^* + q \rightarrow q +Q + \bar Q$ or indirectly to the incoming gluon.
 Actually the
${\cal C}_{k,i}(z,Q^2/M^2,m^2)$ belong to the same class as the hadronic
light parton coefficient functions presented in the first part
of expression (2.6). The only difference is that 
${\cal C}_{k,i}(z,Q^2/M^2,m^2)$
receives contributions from a heavy flavor pair produced
in the final state.

In the third set of terms in (2.6)
the heavy flavor coefficient functions originate
from subprocesses where the virtual photon is attached to one of the
outgoing heavy flavors, as for example
in $\gamma^* + g \rightarrow Q + \bar Q$,
so they are given an additional superscript $H$.  
Finally the fourth set of terms in (2.6) contain the
photonic coefficient functions  
indicated by ${\cal C}_{k,\gamma}$ coming from 
reactions such as $\gamma^* + \gamma \rightarrow q + \bar q$
or $\gamma^* + \gamma \rightarrow Q + \bar Q$. These originate
from hard processes where the (on-shell) real photon is directly
attached to the light or heavy quarks produced in the final state
so there is no need for any convolution integral.

The index $i$ in (2.6) runs over all light active flavors
whose number is given by $n_f$ and $e_i$, $e_H$ stand for the
charges of the light and heavy quarks respectively in units
of $e$. The upper boundary of the integrals in (2.6)
containing the convolution of the parton densities with the
heavy flavor coefficient functions is given by
\begin{equation}
z_{\rm max} = \frac{Q^2}{4m^2+Q^2} \,.
\end{equation}
The parton densities as well as the coefficient functions depend
on the mass factorization scale $M$ except for the ${\cal C}^H_{k,\gamma}$
which can be calculated in pQCD without performing mass factorization.
Notice that in addition to the mass factorization scale $M$ the
quantities in (2.6) also depend on the renormalization scale
$R$ which appears in the pQCD corrections via $\alpha_s(R^2)$.
However in this paper we will put $R$ = $M$.

According to the origin of the photonic parton densities
and the two different types of coefficient functions 
i.e., ${\cal C}_{k,q}$, ${\cal C}_{k,g}$ (hadronic) 
and ${\cal C}_{k,\gamma}$ (photonic) we will 
call the first three terms in (2.6) (represented by the integrals), the  
$\underline{\rm hadronic}$ photon parts, and the last term 
the $\underline{\rm pointlike}$ photon part. Notice
that both these terms are separately factorization scheme 
dependent as indicated by the presence of the
scale $M$. In particular
the scheme dependence of the pointlike photon part in (2.6) is due to the
light quark contribution ${\cal C}_{k,\gamma}(x,Q^2/M^2)$. The 
scheme dependence is cancelled by the hadronic photon part
due to the light quark contribution provided that the
quark densities and the hadronic coefficient functions are 
computed in the same scheme as ${\cal C}_{k,\gamma}(x,Q^2/M^2)$.
 The hadronic heavy flavor
part is scheme dependent in itself. The photonic 
heavy flavor piece is obtained without having to perform
mass factorizaton and needs no parton distribution 
functions, and is thus not dependent on the factorization scheme.

In the subsequent part of this section we will
discuss the contributions to the coefficient functions in (2.6)
which are needed for a next to leading order (NLO) description of the
photon structure functions $F_2^\gamma(x,Q^2)$ and $F_L^\gamma(x,Q^2)$.
The results of our calculations will be presented in the plots of section 3.
For these NLO calculations we also have to use the
next to leading logarithmic (NLL) approximation to the parton
densities, which are  given for example in \cite{grv2},\cite{acfgp},\cite{gs}.

Starting with the NLL parton densities the singlet and
nonsinglet combinations are written in the following way.
Below the charm-quark threshold we have
\begin{equation} 
n_f = 3\quad ,\quad 
\sum_{i=1}^{3}e_i^2 = \frac{2}{3}
\quad ,\quad 
\sum_{i=1}^{3}e_i^4 = \frac{2}{9}\,,
\end{equation}
\begin{equation} 
\Sigma^\gamma = 
u^\gamma + \bar u^\gamma 
+ d^\gamma + \bar d^\gamma 
+ s^\gamma + \bar s^\gamma\,, 
\end{equation}
\begin{equation}
\Delta^\gamma = 
\frac{1}{9}(2u^\gamma + 2\bar u^\gamma 
- d^\gamma - \bar d^\gamma 
- s^\gamma - \bar s^\gamma)\,. 
\end{equation}
Above the charm-quark threshold and below the bottom-quark threshold 
the above quantities are changed into
\begin{equation} 
n_f = 4\quad ,\quad 
\sum_{i=1}^{4}e_i^2 = \frac{10}{9}
\quad ,\quad 
\sum_{i=1}^{4}e_i^4 = \frac{34}{81}\,,
\end{equation}
\begin{equation} 
\Sigma^\gamma = 
u^\gamma + \bar u^\gamma 
+ d^\gamma + \bar d^\gamma 
+ s^\gamma + \bar s^\gamma 
+ c^\gamma + \bar c^\gamma\,, 
\end{equation}
\begin{equation}
\Delta^\gamma = 
\frac{1}{6}(u^\gamma + \bar u^\gamma 
+ c^\gamma + \bar c^\gamma
- d^\gamma - \bar d^\gamma 
- s^\gamma - \bar s^\gamma)\,. 
\end{equation}
Finally above the bottom-quark threshold they become
\begin{equation} 
n_f = 5\quad ,\quad 
\sum_{i=1}^{5}e_i^2 = \frac{11}{9}
\quad ,\quad 
\sum_{i=1}^{5}e_i^4 = \frac{35}{81}\,,
\end{equation}
\begin{equation} 
\Sigma^\gamma = 
u^\gamma + \bar u^\gamma 
+ d^\gamma + \bar d^\gamma 
+ s^\gamma + \bar s^\gamma 
+ c^\gamma + \bar c^\gamma 
+ b^\gamma + \bar b^\gamma\,, 
\end{equation}
\begin{equation}
\Delta^\gamma = 
\frac{1}{15}(3u^\gamma + 3\bar u^\gamma 
+ 3c^\gamma + 3\bar c^\gamma
- 2d^\gamma - 2\bar d^\gamma 
- 2s^\gamma - 2\bar s^\gamma 
- 2b^\gamma - 2\bar b^\gamma)\,. 
\end{equation}
Because the photon is a charge conjugate eigenstate one can put the
quark densities equal to the antiquark densities.

We will now discuss the origin of the coefficient functions
${\cal C}_{k,i}$ $(k=2,L$, $i=q,g,\gamma)$ which appear in (2.6).
Starting with the last terms, the
photonic coefficient functions ${\cal C}_{k,\gamma}$
are given up to next to leading order by the following
parton subprocesses. 
In the Born approximation the light quarks
are produced by the reaction (fig.2)
\begin{equation}
\gamma^*(q) + \gamma(k) \rightarrow q +\bar q\,,
\end{equation}
while the heavy quarks are produced by the same reaction
\begin{equation}
\gamma^*(q) + \gamma(k) \rightarrow Q +\bar Q\,,
\end{equation}
provided the square of the c.m. energy denoted by $s$, where $s=(k+q)^2$,
satisfies the threshold condition $s\ge 4m^2$. 
The $O(\alpha_s)$ pQCD corrections are given by the one-loop contributions
to processes (2.17) and (2.18) (see fig.3) and the gluon bremsstrahlung 
processes (see fig.4)
\begin{equation}
\gamma^*(q) + \gamma(k) \rightarrow q +\bar q +g \,,
\end{equation}
\begin{equation}
\gamma^*(q) + \gamma(k) \rightarrow Q +\bar Q +g \,.
\end{equation}
The parton cross section for the Born reaction in
the case of light quarks (2.17) can be found
in \cite{gr}, \cite{bb}. In the case of heavy flavor production
(2.18) the Born cross section is presented in \cite{dg},
\cite{ggr1}. Notice
that the above reactions are very similar to the ones where the 
on-mass-shell photon $\gamma(k)$ is replaced by a 
gluon $g(k)$. The cross sections of the photon-induced processes 
constitute the Abelian parts of the expressions obtained for the 
gluon-induced processes which are presented up to 
order $\alpha_s^2$ for the case of massless quarks 
in \cite{zn} and in the case of massive quarks in \cite{lrsn1}.
By equating some color factors equal to unity or zero 
in the latter expressions one 
automatically obtains the cross sections for the 
photon-induced processes above in particular
for (2.19) and (2.20) (see Appendix).  In the case of massless 
quarks the parton cross sections for (2.17), (2.19) contain collinear
divergences which can be attributed to the initial photon being 
on-mass-shell. These singularities are removed by mass 
factorization in the following way. We define
\begin{equation}
\hat{\cal F}_{k,\gamma}(z,Q^2,\epsilon) = \sum_i\int^1_0\,dz_1
\int^1_0\,dz_2 \delta(z-z_1z_2) \Gamma_{i\gamma}(z_1,M^2,\epsilon)
\,{\cal C}_{k,i}(z_2,\frac{Q^2}{M^2})\,,
\end{equation}
where $\hat{\cal F}_{k,\gamma}(z,Q^2,\epsilon)$ is the parton 
structure function, which
is related to the parton cross section in the same way as the 
photon structure function $F^\gamma_k(x,Q^2)$ is related to the 
cross section $d^2\sigma/dxdy$ in (2.3).
The parton structure function
contains the collinear divergences represented by the pole terms
$\epsilon^{-j}$ ($j$ is a positive integer) where $\epsilon = n-4$
 (we use dimensional regularization).
These divergences are absorbed in the transition 
functions $\Gamma_{i\gamma}$ $(i=\gamma\,,\,q\,,\,g)$ which depend both on
$\epsilon$ and on the mass factorization scale $M$.
They can be inferred from the Abelian parts of $\Gamma_{ig}$ in
\cite{gr}, \cite{grv1}, \cite{bb} and \cite{fp}.

Both the photonic and hadronic coefficient functions 
${\cal C}_{k,i}$ $(i=\gamma\,,\, q\,,\, g)$ 
which appear in the expressions
for $F_2^\gamma(x,Q^2)$ and $F_L^\gamma(x,Q^2)$ in (2.6) 
are computed in the $\overline{\rm MS}$ scheme. 
The coefficient functions ${\cal C}_{i,k}$ in (2.6) and (2.21)
can be expanded in a power series in $\alpha_s$ as follows
\begin{equation}
{\cal C}_{k,i} =
{\cal C}_{k,i}^{(0)} +
\frac{\alpha_s(M^2)}{4\pi} {\cal C}_{k,i}^{(1)} +
\Big(\frac{\alpha_s(M^2)}{4\pi}\Big)^2 {\cal C}_{k,i}^{(2)} 
+ \cdots 
\end{equation}
which holds for the light as well as the heavy flavor
contributions.  
The photonic coefficient functions
for light quarks ${\cal C}_{k,\gamma}^{(0)}$ and 
${\cal C}_{k,\gamma}^{(1)}$ can be directly derived via the mass 
factorization formula (2.21) from reactions (2.17) and (2.19)
respectively. The heavy flavor coefficients
${\cal C}_{k,\gamma}^{H,(0)}$ and ${\cal C}_{k,\gamma}^{H,(1)}$,
which are obtained without using mass factorization,
originate from processes (2.18) and (2.20). Notice that in
the case of massive quarks the parton structure functions corresponding
to the reactions (2.18) and (2.20) do not have
collinear singularities and they can automatically be identified
with the coefficient functions 
${\cal C}_{k,\gamma}^{H}$. 

Using the mass factorization formula in (2.21) one can
also obtain the order $\alpha_s$ contributions
to the hadronic coefficient functions
${\cal C}_{k,q}^{(1)}$ coming from process (2.19). The higher order
contributions to the hadronic coefficient functions
emerge when one calculates the NLO
corrections to process (2.17). For example the gluonic coefficients
${\cal C}_{k,g}^{(1)}$ can be inferred from the contributions to   
\begin{equation}
\gamma^*(q) + \gamma(k) \rightarrow q + \bar q + q + \bar q\,,
\end{equation}
while ${\cal C}_{k,g}^{H,(1)}$ 
can be inferred from the contributions to 
\begin{equation}
\gamma^*(q) + \gamma(k) \rightarrow q + \bar q + Q + \bar Q\,.
\end{equation}
Fortunately there is a quicker method to obtain the
same information. The hadronic coefficient functions needed for the
 $O(\alpha_s)$ renormalization group improved photon structure
functions $F^\gamma_k(x,Q^2)$ (2.6) can also be obtained from 
deep inelastic lepton hadron scattering,
where the higher order corrections are known. For light flavor 
production we have listed the parton subprocesses and the corresponding 
coefficients which follow from these reactions in table 1.
We have given the corresponding information for heavy flavor
production in table 2. In lowest order the photonic and hadronic
coefficient functions have been presented in the literature (see \cite{gr},
\cite{bb}, \cite{dgr},\cite{ggr1}). Since these authors used a
notation which is different from ours
we will present the relevant formulae below. 
In next to leading order the expressions
for the coefficient functions are obtained from \cite{zn} 
(light quarks and gluons) and \cite{lrsn1} (heavy quarks).
However the expressions are too long to be presented in a paper
\footnote{These functions are available from 
smith@elsebeth.physics.sunysb.edu.}. The method
whereby the higher order coefficients
can be derived from the expressions in \cite{zn}, \cite{lrsn1} is 
explained in the Appendix.

Starting with the photonic coefficients for light quarks
(see reaction (2.17)) they are given by
\begin{equation}
{\cal C}_{2,\gamma}^{(0)}(z,\frac{Q^2}{M^2}) = 
4\{z^2 + (1-z)^2\}\{\ln\frac{Q^2}{M^2} + \ln(1-z) -\ln(z)\}
+32z(1-z) - 4\,,
\end{equation}
and
\begin{equation}
{\cal C}_{L,\gamma}^{(0)}(z,\frac{Q^2}{M^2}) = 
16z(1-z)\,.
\end{equation}
For massive quarks in the final state (see (2.18)) we have
\begin{eqnarray}
{\cal C}_{2,\gamma}^{H,(0)}(z,Q^2,m^2)
&=&
\Big[\Big\{ 4 - 8z(1-z) + \frac{16m^2}{Q^2} z (1-3z)
-  \frac{32m^4}{Q^4} z^2 \Big\} L
\nonumber \\ &&
+ \Big\{ -4 +32z(1-z) - 16\frac{m^2}{Q^2} z(1-z)\Big\}
 \sqrt{1-  \frac{4m^2}{s}}
\Big]\,,\nonumber \\
\end{eqnarray}
and
\begin{eqnarray}
{\cal C}_{L,\gamma}^{H,(0)}(z,Q^2,m^2)
&=&
16z(1-z) \Big[ \sqrt{1-\frac{4m^2}{s}} - 2 \frac{m^2}{s} L\Big]\,,
\end{eqnarray}
where $m$ is the heavy-flavor mass
and $\sqrt{s}$ is the c.m. energy
of the virtual photon-real photon system.
Furthermore we have 
\begin{equation}
s= (1-z)\frac{Q^2}{z} \quad , \quad 
L = \ln\left[\frac{ 1 + \sqrt{1-4m^2/s}}{ 1 - \sqrt{1-4m^2/s}}
\right]\,.
\end{equation}
Formulae (2.27) and (2.28) can be found in \cite{ew1},\cite{mg}.

In the next order in $\alpha_s$ process (2.19) (Fig.4)
and the one-loop corrections to process (2.17) (Fig.3)
give rise to the coefficients $C^{(1)}_{k,\gamma}(z,Q^2/M^2)$.
In the case the outgoing fermion lines in figs.3,4
stand for the heavy flavors (see reactions (2.18) and (2.20))
the corresponding coefficients
are given by ${\cal C}_{k,\gamma}^{H,(1)}(z,Q^2,m^2)$.
More information about the higher order corrections to the
photonic coefficient functions can be found in the Appendix.

In zeroth order of $\alpha_s$ the hadronic coefficient functions are
\begin{eqnarray}
{\cal C}_{2,q}^{(0)}(z,\frac{Q^2}{M^2})&=&  \delta(1-z)\,, \\
{\cal C}_{L,q}^{(0)}(z,\frac{Q^2}{M^2})&=& 0\,, \\
{\cal C}_{k,g}^{(0)}(z,\frac{Q^2}{M^2})&=& 0\,, \qquad {\mbox (k=2,L)}\,. 
\end{eqnarray}
In order $\alpha_S$ the hadronic coefficient functions 
originating from a light quark in the initial state (table 1)
are given by 
\begin{eqnarray}
{\cal C}_{2,q}^{(1)}(z,\frac{Q^2}{M^2})&=& C_F \Big[ \Big\{
\Big(\frac{4}{1-z}\Big)_+ -2 - 2z\Big\} \nonumber \\ &&
\times \Big\{ \ln\frac{Q^2}{M^2} +\ln(1-z) -\frac{3}{4} \Big\}
-2\frac{1+z^2}{1-z} \ln z +\frac{9}{2} +\frac{5}{2} z\nonumber \\ &&
+\delta(1-z)\Big\{3 \ln\frac{Q^2}{M^2}
-9-4\zeta{(2)}\Big\} \Big]\,,
\end{eqnarray}
where the standard definition of a plus distribution
is used, and
\begin{eqnarray}
{\cal C}_{L,q}^{(1)}(z,\frac{Q^2}{M^2})= C_F \Big[ 4z \Big]\,.
\end{eqnarray}
Notice that in order $\alpha_s$ there is no difference
between ${\cal C}_{k,q}^{S,(1)}$ and ${\cal C}_{k,q}^{NS,(1)}$.
The coefficient functions for a gluon in the initial
state and massless quarks in the final state (table 1) can
be derived from (2.25) and (2.26) via multiplication by a color
factor
\begin{equation}
{\cal C}_{k,g}^{(1)}(z,\frac{Q^2}{M^2})
= n_f T_f {\cal C}_{k,\gamma}^{(0)}(z,\frac{Q^2}{M^2})\,, \quad (k=2,L)\,.
\end{equation}
An analogous relation holds when the massless quarks in the
final state are replaced by the heavy flavors (table 2) and we get
from (2.27) and (2.28)
\begin{equation}
{\cal C}_{k,g}^{H,(1)}(z,Q^2,m^2)
= T_f {\cal C}_{k,\gamma}^{(0)}(z,Q^2,m^2)\,, \quad (k=2,L)\,.
\end{equation}
The color factors which appear in the above equations are given
by $C_F=4/3$ and $T_f = 1/2$ for the case of $SU(3)$. 

The higher order $\alpha_s^2$ corrections to the 
coefficient functions, describing massless partons only, are denoted by
${\cal C}_{k,i}^{(2)}$ where $i=q,g$ (see table 1).
They have been calculated in \cite{zn}.
In the Appendix we have decomposed ${\cal C}_{k,i}^{(2)}$ into color
factors so that we can infer the $O(\alpha_s)$ photonic
coefficients ${\cal C}_{k,\gamma}^{(1)}$ 
from the Abelian part of ${\cal C}_{k,g}^{(2)}$.

The $O(\alpha_s^2)$ corrections to the heavy
flavor coefficient functions
given by ${\cal C}_{k,i}^{(2)}(z,Q^2/M^2,m^2)$ 
and ${\cal C}_{k,i}^{H,(2)}(z,Q^2/M^2,m^2)$ 
(table 2) are calculated for the first time in \cite{lrsn1}. The 
relations between these coefficients and the ones derived in 
section 5 of \cite{lrsn1} will be presented in the Appendix.
By decomposing them in color factors we again can derive the
photonic heavy flavor coefficient ${\cal C}_{k,\gamma}^{H,(1)}$ from the
Abelian part of ${\cal C}_{k,g}^{H,(2)}$. Since in lowest order
the hadronic heavy flavor coefficient ${\cal C}_{k,i}^{(2)}(z,Q^2/M^2,m^2)$
only contributes up to the $O(\alpha_s^2)$ level,
when $i=q$ we do not have to distinguish between singlet
(S) and non-singlet (NS)
and we can put
\begin{equation}
{\cal C}_{k,q}^{S,(2)}(z,\frac{Q^2}{M^2},m^2)
= {\cal C}_{k,q}^{NS,(2)}(z,\frac{Q^2}{M^2},m^2) 
= {\cal C}_{k,q}^{(2)}(z,Q^2,m^2) \,.
\end{equation} 
The above expression indicates that in lowest order
${\cal C}_{k,q}^{(2)}(z,Q^2,m^2)$ is determined without having performed
mass factorization which is indicated by its independence of
the mass factorization scale $M$. This is because it originates from
the Compton scattering process, which in lowest order does not have
collinear singularities.

Finally in table 3 we have translated our notations for the
coefficient functions into those used in \cite{gr}, \cite{dgr}, 
\cite{ggr1}, \cite{bb}.
We also list the new contributions to the
photon structure functions which were not included 
earlier in the literature.

\newpage
\mysection{Results}
In this section we will discuss the 
NLO QCD corrections to the photon structure functions
$F^\gamma_k(x,Q^2)$ for $k=2,L$.
In particular we focus our attention on the heavy flavor contributions
(mainly charm), which originate from the hadronic as well as the photonic
coefficient functions in (2.6).  Since heavy flavors can be 
produced either in virtual-photon parton or in virtual-photon 
real-photon reactions we will call the former hadronic
heavy flavor production and the latter photonic heavy flavor production.

In the subsequent part of this section we want to make a comparison
between the LO and NLO description of the photon structure functions, where all
contributions listed in tables 4 and 5 are included.
Furthermore we want to investigate the relative magnitude
of the heavy flavor (mainly charm) component of the structure function.
We also show the difference
between the massless and massive heavy flavor approach.
When the heavy quarks are treated as massless,
their contribution to the photon structure functions are
given by the corresponding parton densities in the photon convoluted
with the light quark and gluon coefficient functions.
This description is appropriate when $Q^2 >> m^2$.
If $Q^2$ is of the same order of magnitude as $m^2$, then the massive
quark approach has to be adopted and the heavy flavor production
is described by the heavy flavor coefficient functions in (2.6)
which can be computed order-by-order in perturbation theory.

In the literature a LO analysis was given
for $F^\gamma_2(x,Q^2)$ in \cite{dg} and $F^\gamma_L(x,Q^2)$ in \cite{dgr}. 
Here all LO coefficient functions in tables 4 and 5 were included except for
the ones related to hadronic heavy flavor production,
(i.e., $ \gamma^* + g \rightarrow Q + \bar Q$).
The last contributions were also neglected in the NLO 
analysis for $F^\gamma_2(x,Q^2)$ in \cite{grv2} and the photonic heavy 
flavor contribution from $\gamma^* + \gamma \rightarrow Q + \bar Q$
was only taken into account in lowest order. A NLO analysis of 
$F^\gamma_L(x,Q^2)$
could not be carried out previously because the
order $\alpha_s^2$ contributions to all the longitudinal coefficient
functions were not known until recently. Since all NLO coefficient
functions are now known, and they are listed in tables 4 and 5, we are able
to present a complete NLO description for both $F^\gamma_2(x,Q^2)$ 
and for $F^\gamma_L(x,Q^2)$ as well as
make a comparison with the LO descriptions.

In our plots we adopt the LO and NLO parametrizations of the parton
densities in the photon from \cite{grv2} 
(for other sets see \cite{acfgp}, \cite{gs}). 
For $n_f=3$ we use $\Lambda_{QCD} = 232$
MeV at leading order and $\Lambda_{QCD} = 248$ MeV at next to leading
order.  For $n_f = 4$, both the leading order and the next
to leading order $\Lambda_{QCD}$ are set equal to 200 MeV.
In leading order, we use a one-loop result for the running coupling constant
and in next to leading order a two-loop corrected running coupling constant 
is chosen,
see e.g. \cite{grv2}. All calculations are done with $M^2 = Q^2$, 
except where otherwise indicated.
In our analysis, when we take the charm quark to be massive, we take
$m_c=1.5 \, {\rm GeV}/c^2$.  
Furthermore we take three light flavors ($n_f = 3$) for the parton
densities, the coefficient functions and the running coupling constant.
When we treat the charmed quark as massless, it then takes
on the identity of an ordinary parton, so we set $n_f = 4$.
The bottom and top quark contributions will be omitted since they 
are negligible
for the $Q^2$ values accessible at past and present experiments.
In the LO approximation the corresponding parton densities are 
multiplied by the
coefficient functions in tables 4 and 5, which are indicated by LO.
In NLO we have chosen the $\overline{\rm MS}$ scheme for the parton
densities, the coefficient functions and the running coupling constant.
The coefficient  functions which have to be added to the LO ones are
indicated by NLO in tables 4 and 5. In order to get a consistent
NLO analysis for the structure functions we follow the procedure
in \cite{grv2}, which is explained in \cite{grv1}.
Therefore we multiply the LO coefficient functions by $f^\gamma$
and the NLO coefficient functions by
$f^\gamma_o$ in (2.6) (for the notation of $f^\gamma$ and $f^\gamma_o$ 
see eqn. (A.23) and the discussion in the Appendix A in \cite{grv2}).
Notice that in \cite{grv2} the parton densities described in Appendix A
were presented in the DIS$_\gamma$ scheme. However they can be
changed into the $\overline{\rm MS}$ scheme via eqns.(4)-(6) 
in \cite{grv2}.
After changing the lowest order photonic
coefficient function $C^{(0)}_{2,\gamma}$
in the DIS$_\gamma$ scheme we have checked that
both schemes lead to the same result
provided the change of eqn.(4) in \cite{grv1}
is only applied to the parton density denoted by $f^\gamma$
as defined above.

We now compare the results from our calculations for $F_2^\gamma(x,Q^2)$
first with data from PLUTO \cite{pluto} 
($Q^2 = 5.9$ (GeV/$c$)$^2$) and then
with data from AMY \cite{amy} ($Q^2 = 51$ (GeV/$c$)$^2$).
We also show predictions for $F_L^\gamma(x,Q^2)$.

In fig.5 we make a comparison between the LO and NLO
aproximation for 
$F^\gamma_2(x,Q^2)$ at $Q^2=5.9 \, ({\rm GeV}/c)^2$, where the 
heavy charm components (hadronic and photonic) are included. 
The low-$x$ hump is due to charm production, which turns off 
at about $x = 0.4$ (the threshold value). We also show separately
the contributions due to massive charm production. When this 
contribution reaches its maximum value it constitutes about
20 \% of the structure function $F^\gamma_2$ in LO and
30 \% in NLO. The $O(\alpha_S)$ correction to the Born
contributions to massive charm production
are quite large, adding approximately another
50 \% to the Born terms.
Overall, we observe that LO
and NLO are not very different. Note that the data also seem
to indicate the presence of a charm component.

In fig.6 we do the same for $F^\gamma_L(x,Q^2)$ at $Q^2=5.9\,({\rm GeV}/c)^2$.
This is for theoretical purposes only: there are no data presently
available for $F^\gamma_L$ at any value of $Q^2$.
We see from this and the previous figure that there is not much
difference between the LO and NLO results both 
for $F_2^\gamma$ and $F_L^\gamma$.
However, the heavy charm component of $F_L^\gamma$ is less important 
than in the case of $F^\gamma_2$. At LO it is about 15 \%
where this component reaches its maximum, whereas in NLO
it amounts to about about 30 \% also. The latter is due
to the fact that the $O(\alpha_S)$ corrections to 
the heavy charm component of $F_L^\gamma$ 
are as large as  100 \%.

In fig.7 we present $F^\gamma_2(x,Q^2)$ at LO for three different
choices of mass factorization scale. Note that in this case the
only variation is due to the parton densities. The variation in the 
$M$ dependence is uniform over the whole $x$-range.
In fig.8 we do the same for $F^\gamma_L(x,Q^2)$ at $Q^2=5.9\,({\rm GeV}/c)^2$.
Here there is additional scale dependence due to $\alpha_s(M^2)$.
Hence, contrary to fig.7, the curve for $M=Q/2$ is the upper one here.

Fig.9 shows the same as fig.7 but now at NLO. There is now additional
scale dependence due to $\alpha_s(M^2)$ and the mass factorization scale
logarithms of the type $\ln(Q^2/M^2)$
in the coefficient functions
(see e.g (2.25) and (2.33)). Note that the scale
dependence is reduced in the small-$x$ region
compared to the LO case. However at very large
$x$ values, where the charm contribution can
be neglected, the scale variation
is larger than in the LO case. 
This is due to the pointlike light quark contribution,
which drops increasingly dramatically as one increases
$M$.  At small $x$ this is partially offset by the increase of the charm
contribution.

In fig.10 we show the same plots as in fig.9 for $F^\gamma_L(x,Q^2)$.
The scale variation is small as in the LO case.

We now turn to a comparison of results for massive versus massless
charm contributions as defined above. Since the differences
are essentially the same in the LO case as in the NLO case
we only show plots for the latter. Therefore 
in fig.11 we compare the NLO massless ($n_f=3$) plus the 
massive charm-quark contribution to
$F^\gamma_2(x,Q^2)$ at $Q^2=5.9 \, ({\rm GeV}/c)^2$, 
with the NLO massless ($n_f=4$) contribution.
Note that the massless $n_f=4$ contribution is smaller than
the curve where we take $n_f=3$ massless and a massive charm quark, 
even at large $x$ where
the charm contribution is zero. This is due to a change
in $\Lambda_{QCD}$ and consequently a
 change in the parton distribution functions.
However the difference between the massless and massive cases is small
for large $x$, where threshold effects are negligible. 

In fig.12 we show the same plots for $F^\gamma_L(x,Q^2)$ in NLO
at $Q^2=5.9\,({\rm GeV}/c)^2$. Note the enormous increase that
occurs in going to the $n_f=4$ massless case. 
Since this effect is already there in the
LO case it can be understood as follows. 
In the case of $n_f=3$ where charm is considered massive one includes
the coefficient function ${\cal C}^{(0)}_{L,\gamma}$ (2.26),
which is multiplied by $2/9$ and ${\cal C}^{H,(0)}_{L,\gamma}$
(2.28), which is multiplied by $16/81$.
If $n_f=4$ the charm is treated as massless and
${\cal C}^{H,(0)}_{L,\gamma}$ (massive charm) is replaced by
${\cal C}^{(0)}_{L,\gamma}$ (massless charm).
Since  the latter is much larger than the former
due to the additional suppression factor
in (2.28) this explains why the result for $n_f=4$
is much larger than for $n_f=3$.

In fig.13 we show the $x$-dependences of the massive hadronic charm 
contribution and the massive photonic charm contribution to
$F^\gamma_2(x,Q^2)$ at $Q^2=5.9 \, ({\rm GeV}/c)^2$ in LO and
in NLO. The corresponding results for $F^\gamma_L(x,Q^2)$
are shown in fig.14 in LO and in NLO. The interesting feature
to note in all these figures is the complete dominance of the photonic
charm production over the hadronic production.  
This makes $F^\gamma_2(x,Q^2)$ for massive charm production at moderate
$x$ a very promising test of pQCD, because
of the lack of dependence on the hadronic component. Experimentally
this is of course a very difficult quantity to determine, but
perhaps not impossible. The same holds for $F^\gamma_L(x,Q^2)$
for massive charm production, but that is even more difficult
to determine experimentally.
However for $x<0.01$ the pointlike 
contributions to both $F^\gamma_2$ and
$F^\gamma_L$ for massive charm production become very small
and the hadronic component begins to dominate.

We now repeat all the figures for the $Q^2 = 51\, ({\rm GeV}/c)^2$
value of the AMY collaboration. We remark that now the charm contribution
switches off at $x = 0.85$. 
Here the heavy charm component becomes in general larger
than in the case for $Q^2=5.9\,({\rm GeV}/c)^2$. For
$F_2^\gamma$ it is 30 \% in LO where this component reaches
its maximum,
and 40 \% in NLO.
For $F_L^\gamma$ the percentages are roughly similar. Note
however that the $O(\alpha_s)$ corrections are smaller
than for $Q^2=5.9\,({\rm GeV}/c)^2$. For $F_2^\gamma$ 
they are up to 15 \% and for $F_L^\gamma$ up to 30 \%.
The mass factorization scale dependence
at large $x$ for $F^\gamma_2(x,Q^2)$ at NLO seems (fig.19)
to be somewhat reduced compared to the case of
$Q^2 = 5.9\, ({\rm GeV}/c)^2$
but still larger than at LO (fig.17).

To conclude, we have presented in this paper the
first complete NLO analysis of $F^\gamma_2(x,Q^2)$ and $F^\gamma_L(x,Q^2)$
containing both light and heavy quarks. Summarizing our findings we have
seen that for both values of $Q^2$ we considered the NLO structure 
functions are not too different from the LO ones. This is not 
so surprising for $F^\gamma_2(x,Q^2)$ since we used the parton densities of 
\cite{grv2} and most of the contributions
were already included in their analysis except for 
$O(\alpha_S)$ corrections to heavy quark production,
which are numerically small. 
We see that $F^\gamma_2 $ has a moderate
sensitivity to changes in the mass factorization scale except at large $x$.

For $F^\gamma_L(x,Q^2)$ this is the first NLO analysis, and at the 
same time complete, since all heavy and light quark contributions
have been included. We found that $F^\gamma_L(x,Q^2)$ changes very
little from LO to NLO, and is very stable under scale changes. 
Above $x\approx 0.1$ the hadronic production
of charm is small compared with the photonic production,
while the former is dominant for $x< 0.01$.
All this would make a measurement of $F^\gamma_L(x,Q^2)$
(e.g. at LEP2) an interesting prospect.

Our results could be used to determine  more
accurate NLL parton distribution functions for the
photon. This would become especially relevant when 
data become available for $F^\gamma_2$ for charm
production, and for $F^\gamma_L$.
Finally, we stress
that if the heavy quark contribution could be extracted from 
a measurement of $F^\gamma_2$ this would
yield a very good test of perturbative QCD.

{\bf Acknowledgements}

The work in this paper was supported in part under the
contracts NSF 92-11367 and DOE DE-AC02-76CH03000. 
Financial support was also provided by the Texas National
Research Laboratory Commission. S.R. would like to thank Fermi National
Accelerator Laboratory for their hospitality while this paper was being 
completed.

\vfill
\newpage
%
\newcommand{\ra}{\gamma^*+q(\bar q) \rightarrow q(\bar q)}
\newcommand{\rb}{\gamma^*+q(\bar q) \rightarrow q(\bar q)+g}
\newcommand{\rc}{\gamma^*+g \rightarrow q+\bar q}
\newcommand{\rd}{\gamma^*+q(\bar q) \rightarrow q(\bar q)+g+g}
\newcommand{\re}{\gamma^*+q(\bar q) \rightarrow q(\bar q)+q(\bar q)+\bar q(q)}
\newcommand{\rf}{\gamma^*+g \rightarrow q+\bar q+g}
\newcommand{\ckq}{{\cal C}_{k,q}^{(0)}}
\newcommand{\ckqSa}{{\cal C}_{k,q}^{S,(1)}}
\newcommand{\ckqNSa}{{\cal C}_{k,q}^{NS,(1)}}
\newcommand{\ckqSb}{{\cal C}_{k,q}^{S,(2)}}
\newcommand{\ckqNSb}{{\cal C}_{k,q}^{NS,(2)}}
\newcommand{\ckga}{{\cal C}_{k,g}^{(1)}}
\newcommand{\ckgb}{{\cal C}_{k,g}^{(2)}}

\centerline{\bf \large{Table 1.}}
\vspace{2cm}

\hspace{-0.1cm}\begin{tabular}{||c||c|c||} \hline 
{\rm order} & {\rm parton subprocess} & {\rm coefficient function} \\ \hline

$\alpha_s^0$ & $\ra$ & $\ckq$           \\   
${}$         & ${}$  &   ${}$            \\ \hline
$\alpha_s^1$ & $\rb$ & $\ckqNSa = \ckqSa$   \\   
${}$         & ${}$  &   ${}$            \\ \hline
${}$         & $\rc$ & $\ckga$          \\   
${}$         & ${}$  &   ${}$            \\ \hline
$\alpha_s^2$ & $\rd$ & $\ckqNSb = \ckqSb$ \\   
${}$         & ${}$  &   ${}$            \\ \hline
${}$         & $\re$ & $\ckqNSb \ne \ckqSb$  \\   
${}$         & ${}$  &   ${}$            \\ \hline
${}$         & $\rf$ & $\ckgb$            \\  
${}$         & ${}$  &   ${}$            \\ \hline
\end{tabular}

\vspace{0.5cm}

List of deep inelastic virtual-photon-parton subprocesses up to
$O(\alpha_s^2)$. The one and two-loop corrections to the lower
order processes have been included in our calculations
but are not explicitly mentioned in the table.

%

%
\newcommand{\rg}{\gamma^*+g \rightarrow Q + \bar Q}
\newcommand{\rh}{\gamma^*+g \rightarrow Q + \bar Q+g}
\newcommand{\ri}{\gamma^*+q(\bar q) \rightarrow q(\bar q)+Q+\bar Q}
\newcommand{\ckgHa}{C_{k,g}^{H,(1)}}
\newcommand{\ckgHb}{C_{k,g}^{H,(2)}}
\newcommand{\ckqHb}{C_{k,q}^{H,(2)}}

\centerline{\bf \large{Table 2.}}
\vspace{2cm}

\hspace{-0.1cm}\begin{tabular}{||c||c|c||} \hline 
{\rm order} & {\rm parton subprocess} & {\rm coefficient function} \\ \hline

$\alpha_s^1$ & $\rg$ & $\ckgHa$   \\   
${}$         & ${}$  & ${}$       \\ \hline
$\alpha_s^2$ & $\rh$ & $\ckgHb$   \\   
${}$         & ${}$  & ${}$       \\ \hline
$\alpha_s^2$ & $\ri$ & $\ckqHb\,,\, \ckqNSb= \ckqSb$ \\   
${}$         & ${}$  & ${}$       \\ \hline
\end{tabular}

\vspace{0.5cm}

List of deep inelastic virtual-photon-partonic subprocesses 
contributing to heavy flavour production up to
$O(\alpha_s^2)$. The one-loop corrections to the Born
approximation have been included in our calculations
but are not explicitly mentioned in the table.

%

%
\newcommand{\cca}{{\cal C}_{2,\gamma}^{(0)} \,\, (2.25)}
\newcommand{\ccb}{\frac{3\alpha_s}{4\pi} e_q^4
{\cal C}_{L,\gamma}^{(0)} \,\, (2.26)}
\newcommand{\ccc}{\frac{3\alpha_s}{4\pi}(\frac{2}{3})^4
{\cal C}_{2,\gamma}^{H,(0)} \,\, (2.27)}
\newcommand{\ccd}{\frac{3\alpha_s}{4\pi}(\frac{2}{3})^4
{\cal C}_{L,\gamma}^{H,(0)} \,\, (2.28)}
\newcommand{\cce}{{\cal C}_{2,q}^{(1)} \,\, (2.33)}
\newcommand{\ccf}{{\cal C}_{L,q}^{(1)} \,\, (2.34)}
\newcommand{\ccg}{{\cal C}_{2,g}^{(1)} \,\, (2.35)}
\newcommand{\cch}{{\cal C}_{L,g}^{(1)} \,\, (2.35)}
\newcommand{\cci}{{\cal C}_{k,g}^{H,(1)} \,\, (2.36)}
\newcommand{\ccj}{{\cal C}_{k,\gamma}^{(1)} \,\, \cite{zn}}
\newcommand{\cck}{{\cal C}_{k,\gamma}^{H,(1)} \,\, \cite{lrsn1}}
\newcommand{\ccl}{{\cal C}_{k,q}^{(2)} \,\, \cite{zn}}
\newcommand{\ccm}{{\cal C}_{k,q}^{H,(2)} \,\, \cite{lrsn1}}
\newcommand{\bba}{B^{(n)}_{\gamma} \,\, (4.12)}
\newcommand{\bbb}{B^{(n)}_{\rm NS}, B^{(n)}_\psi \,\, (4.10)}
\newcommand{\bbc}{B^{(n)}_{\rm G} \,\, (4.11)}
\newcommand{\bbd}{B_{\gamma} \,\, (3.7)}
\newcommand{\bbe}{B_{\rm NS}, B_q \,\, (3.7)}
\newcommand{\bbf}{B_{\rm G} \,\, (3.7)}
\newcommand{\ffa}{\frac{1}{x}F^{\gamma,(0)}_{L,q\bar q} \,\, (15)^{*}}
\newcommand{\ffb}{\frac{1}{x}F^\gamma_{2,c} \,\, (2.13)^{**}}
\newcommand{\ffc}{\frac{1}{x}F^{\gamma,(0)}_{L,q\bar q} \,\, (16)^{*}}
\centerline{\bf \large{Table 3.}}
\vspace{1cm}

\hspace{-0.1cm}\begin{tabular}{||c||c|c|c||} \hline 
{\rm this paper} & \cite{bb} & \cite{gr} & \cite{dgr}${}^*$ 
\,, \cite{ggr1}${}^{**}$ \\ \hline

$\cca $ & $\bba$ & $\bbd$ & ${}$   \\   
${}$    & ${}$   & ${}$   & ${}$ \\ \hline 
$\ccb $ & ${}$   & ${}$   & $\ffa$  \\   
${}$    & ${}$   & ${}$   & ${}$ \\ \hline 
$\ccc $ & ${}$   & ${}$   & $\ffb$ \\   
${}$    & ${}$   & ${}$   & ${}$ \\ \hline 
$\ccd $ & ${}$   & ${}$   & $\ffc$ \\   
${}$    & ${}$   & ${}$   & ${}$ \\ \hline 
$\cce $ & $\bbb$ & $\bbe$ & ${}$   \\   
${}$    & ${}$   & ${}$   & ${}$ \\ \hline 
$\ccf $ & ${}$   & ${}$   & ${}$  \\   
${}$    & ${}$   & ${}$   & ${}$ \\ \hline 
$\ccg $ & $\bbc$ & $\bbf$ & ${}$   \\   
${}$    & ${}$   & ${}$   & ${}$ \\ \hline 
$\cch $ & ${}$   & ${}$   & ${}$   \\   
${}$    & ${}$   & ${}$   & ${}$ \\ \hline 
$\cci $ & ${}$   & ${}$   & ${}$   \\   
${}$    & ${}$   & ${}$   & ${}$ \\ \hline 
$\ccj $ & ${}$   & ${}$   & ${}$   \\   
${}$    & ${}$   & ${}$   & ${}$ \\ \hline 
$\cck $ & ${}$   & ${}$   & ${}$   \\   
${}$    & ${}$   & ${}$   & ${}$ \\ \hline 
$\ccl $ & ${}$   & ${}$   & ${}$   \\   
${}$    & ${}$   & ${}$   & ${}$ \\ \hline 
$\ccm $ & ${}$   & ${}$   & ${}$   \\   
${}$    & ${}$   & ${}$   & ${}$ \\ \hline 
\end{tabular}

\vspace{0.5cm}

Notations in several papers for the hadronic and photonic
coefficient functions. Notice that the expressions in 
\cite{bb} are in Mellin transform space. The blanks mean 
that these contributions were not considered in the papers quoted.

%

%
\newcommand{\rj}{\gamma^*+\gamma \rightarrow q + \bar q}
\newcommand{\rk}{\gamma^*+\gamma \rightarrow Q + \bar Q}
\newcommand{\rl}{\gamma^*+\gamma \rightarrow Q + \bar Q + g}
\newcommand{\eaa}{{\cal C}_{2,q}^{(0)}}
\newcommand{\eab}{{\cal C}_{2,\gamma}^{(0)}}
\newcommand{\eac}{{\cal C}_{2,\gamma}^{H,(0)}}
\newcommand{\ead}{{\cal C}_{2,q}^{NS,(1)}(={\cal C}_{2,q}^{S,(1)})}
\newcommand{\eae}{{\cal C}_{2,g}^{(1)}}
\newcommand{\eaf}{{\cal C}_{2,g}^{H,(1)}}
\newcommand{\eag}{{\cal C}_{2,\gamma}^{H,(1)}}
\newcommand{\eah}{{\cal C}_{2,g}^{H,(1)}}
\newcommand{\eai}{{\cal C}_{2,q}^{H,(2)}}
\newcommand{\eaj}{{\cal C}_{2,q}^{NS,(2)}}
\newcommand{\eak}{{\cal C}_{2,q}^{S,(2)}}

\centerline{\bf \large{Table 4.}}
\vspace{2cm}

\hspace{-0.1cm}\begin{tabular}{||c||c|c||} \hline 
{\rm order} & {\rm parton subprocess} & {\rm coefficient function} \\ \hline

$\alpha_s^0$ & $\ra$ & $\eaa$ \,\,LO, \,\,NLO           \\   
${}$         & ${}$  &   ${}$            \\ \hline
${}$         & $\rj$ & $\eab$ \,\,NLO   \\   
${}$         & ${}$  &   ${}$            \\ \hline
${}$         & $\rk$ & $\eac$ \,\, LO,\,\, NLO        \\   
${}$         & ${}$  &   ${}$            \\ \hline
$\alpha_s^1$ & $\rb$ & $\ead$ \,\, NLO\\   
${}$         & ${}$  &   ${}$            \\ \hline
${}$         & $\rc$ & $\eae $  \,\, NLO\\   
${}$         & ${}$  &   ${}$            \\ \hline
${}$         & $\rg$ & $\eaf$   \,\,LO,\,\,NLO         \\  
${}$         & ${}$  &   ${}$            \\ \hline
${}$         & $\rl$ & $\eag$   \,\,NLO         \\  
${}$         & ${}$  &   ${}$            \\ \hline
$\alpha_s^2$ & $\rh$ & $\eah$ \,\, NLO\\   
${}$         & ${}$  &   ${}$            \\ \hline
${}$         & $\ri$ & $\eai\,,\eaj (= \eak)$  \,\, NLO\\   
${}$         & ${}$  &   ${}$            \\ \hline
\end{tabular}

\vspace{0.5cm}
Coefficient functions used in this paper for a leading
order (LO) and a next-to-leading order (NLO) analysis
of $F_2^\gamma(x,Q^2)/\alpha$.

%

%
\newcommand{\rn}{\gamma^* +\gamma \rightarrow q + \bar q +g}
\newcommand{\ro}{\gamma^* +q(\bar q)\rightarrow q(\bar q) + g+g}
\newcommand{\rp}{\gamma^* +q(\bar q)\rightarrow q(\bar q)+q(\bar q)+\bar q(q)}
\newcommand{\gaa}{{\cal C}_{L,\gamma}^{(0)}}
\newcommand{\gab}{{\cal C}_{L,\gamma}^{H,(0)}}
\newcommand{\gac}{{\cal C}_{L,q}^{NS,(1)}(={\cal C}_{L,q}^{S,(1)})}
\newcommand{\gad}{{\cal C}_{L,g}^{(1)}}
\newcommand{\gae}{{\cal C}_{L,g}^{H,(1)}}
\newcommand{\gaf}{{\cal C}_{L,\gamma}^{(1)}}
\newcommand{\gag}{{\cal C}_{L,\gamma}^{H,(1)}}
\newcommand{\gah}{{\cal C}_{L,q}^{NS,(2)}(={\cal C}_{L,q}^{S,(2)})}
\newcommand{\gai}{{\cal C}_{L,q}^{NS,(2)}\,,{\cal C}_{L,q}^{S,(2)} (\ne{\cal C}_{L,q}^{NS,(2)})}
\newcommand{\gaj}{{\cal C}_{L,g}^{(2)}}
\newcommand{\gak}{{\cal C}_{L,g}^{H,(2)}}
\newcommand{\gal}{{\cal C}_{L,q}^{H,(2)}}
\newcommand{\gam}{{\cal C}_{L,q}^{NS,(2)}(={\cal C}_{L,q}^{S,(2)})}

\centerline{\bf \large{Table 5.}}
\vspace{2cm}

\hspace{-0.1cm}\begin{tabular}{||c||c|c||} \hline 
{\rm order} & {\rm parton subprocess} & {\rm coefficient function} \\ \hline

$\alpha_s^0$ & $\rj$ & $\gaa$ \,\,LO, \,\,NLO           \\   
${}$         & ${}$  &   ${}$            \\ \hline
${}$         & $\rk$ & $\gab$ \,\,LO,\,\,NLO   \\   
${}$         & ${}$  &   ${}$            \\ \hline
$\alpha_s^1$ & $\rb$ & $\gac$ \,\, LO,\,\, NLO        \\   
${}$         & ${}$  &   ${}$            \\ \hline
${}$         & $\rc$ & $\gad$ \,\, LO,\,\,NLO\\   
${}$         & ${}$  &   ${}$            \\ \hline
${}$         & $\rg$ & $\gae $  \,\, LO,\,\,NLO\\   
${}$         & ${}$  &   ${}$            \\ \hline
${}$         & $\rn$ & $\gaf$   \,\,NLO         \\  
${}$         & ${}$  &   ${}$            \\ \hline
${}$         & $\rl$ & $\gag$   \,\,NLO         \\  
${}$         & ${}$  &   ${}$            \\ \hline
$\alpha_s^2$ & $\ro$ & $\gah $ \,\, NLO\\   
${}$         & ${}$  &   ${}$            \\ \hline
${}$         & $\rp$ & $\gai $  \,\, NLO\\   
${}$         & ${}$  &   ${}$            \\ \hline
${}$         & $\rf$ & $\gaj $  \,\, NLO\\   
${}$         & ${}$  &   ${}$            \\ \hline
${}$         & $\rh$ & $\gak $  \,\, NLO\\   
${}$         & ${}$  &   ${}$            \\ \hline
${}$         & $\ri$ & $\gal\,,\gam $  \,\, NLO\\   
${}$         & ${}$  &   ${}$            \\ \hline
\end{tabular}

\vspace{0.5cm}
Coefficient functions used in this paper for a leading
order (LO) and a next-to-leading order (NLO) analysis
of $F_L^\gamma(x,Q^2)/\alpha$.

%

\appendix
\mysection*{Appendix}
\setcounter{section}{1}
In this Appendix we show how one can derive the $O(\alpha_s^2)$
coefficients corresponding to the reactions in tables 1 and 2
from the expressions calculated in \cite{zn} and \cite{lrsn1}
respectively. The $O(\alpha_s^2)$ coefficients mentioned in 
table 1 are given by
\begin{eqnarray}
{\cal C}_{k,q}^{NS,(2)}(z,\frac{Q^2}{M^2})
&=& C_F^2 B_{FF}^{(k)}(z,\frac{Q^2}{M^2})
+ C_A C_F B_{AF}^{(k)}(z,\frac{Q^2}{M^2})
\nonumber \\ &&
+ \qquad n_fT_F C_F B_{FF}^{(k)}(z,\frac{Q^2}{M^2}) \,,
\end{eqnarray}
where ${\cal C}_{L,q}^{NS,(2)}$ and ${\cal C}_{2,q}^{NS,(2)}$ 
are the coefficients
of the $(\alpha_s/4\pi)^2$ term in eqns.(B.1) and (B.2) of \cite{zn}
respectively.
The singlet coefficients can be split into a nonsinglet and a
pure singlet piece as follows
\begin{eqnarray}
{\cal C}_{k,q}^{S,(2)}(z,\frac{Q^2}{M^2})
= {\cal C}_{k,q}^{NS,(2)}(z,\frac{Q^2}{M^2})
+ {\cal C}_{k,q}^{PS,(2)}(z,\frac{Q^2}{M^2}) \,.
\end{eqnarray}
The pure singlet coefficients ${\cal C}_{k,q}^{PS,(2)}$ can be written as
\begin{eqnarray}
{\cal C}_{k,q}^{PS,(2)}(z,\frac{Q^2}{M^2})
= n_f T_f C_F D_{FF}^{(k)}(z,\frac{Q^2}{M^2})\,,
\end{eqnarray}
where ${\cal C}_{L,q}^{PS,(2)}$ and ${\cal C}_{2,q}^{PS,(2)}$ 
are the coefficients
of the $(\alpha_s/4\pi)^2$ terms in eqns.(B.3) and (B.4) of \cite{zn}
respectively. Finally the gluonic coefficient is given by
\begin{eqnarray}
{\cal C}_{k,g}^{(2)}(z,\frac{Q^2}{M^2})
= n_f T_f C_F E_{FF}^{(k)}(z,\frac{Q^2}{M^2})
+ n_fT_f C_A E_{FA}^{(k)}(z,\frac{Q^2}{M^2}) \,,
\end{eqnarray}
where ${\cal C}_{L,g}^{(2)}$ and ${\cal C}_{2,g}^{(2)}$ 
are the coefficients
of the $(\alpha_s/4\pi)^2$ terms in eqns.(B.5) and (B.6) of \cite{zn}
respectively. The color factors in $SU(3)$ are given by $C_F=4/3$, 
$C_A=3$, $T_F=1/2$ and $n_f$ denotes the number of light flavors. 
The $O(\alpha_s)$ photonic coefficient ${\cal C}_{k,\gamma}^{(1)}$
can be derived from the Abelian part of
${\cal C}_{k,g}^{(2)}$ (A.4) and it equals
\begin{eqnarray}
{\cal C}_{k,\gamma}^{(1)}(z,\frac{Q^2}{M^2}) = 
C_FE_{FF}^{(k)}(z,\frac{Q^2}{M^2})\,.
\end{eqnarray}
The coefficient functions due to heavy flavor production (see table 2)
are related to the coefficients defined in \cite{lrsn1} in the
following way. In first order in $\alpha_s$ we have (see also (2.36))
\begin{eqnarray}
{\cal C}_{L,g}^{H,(1)}(z,Q^2,m^2) = \frac{1}{\pi}\frac{Q^2}{m^2z}
 c_{L,g}^{(0)}(\eta,\xi) \,,
\end{eqnarray}
\begin{eqnarray}
{\cal C}_{2,g}^{H,(1)}(z,Q^2,m^2) = \frac{1}{\pi}\frac{Q^2}{m^2z}
\{c_{T,g}^{(0)}(\eta,\xi) + c_{L,g}^{(0)}(\eta,\xi)\}\,,
\end{eqnarray}
with
\begin{eqnarray}
\eta = \frac{s}{4m^2}-1 \quad ,\quad \xi = \frac{Q^2}{m^2}\,.
\end{eqnarray}
In second order in $\alpha_s$ one gets for $i=q,g$
\begin{eqnarray}
{\cal C}_{L,q}^{(2)}(z,Q^2,m^2) = 16\pi\frac{Q^2}{m^2z}
d_{L,q}^{(1)}(\eta,\xi) \,,
\end{eqnarray}
\begin{eqnarray}
{\cal C}_{2,g}^{(2)}(z,Q^2,m^2) = 16\pi\frac{Q^2}{m^2z}
\{d_{T,q}^{(1)}(\eta,\xi) + d_{L,q}^{(1)}(\eta,\xi)\}\,,
\end{eqnarray}
and
\begin{eqnarray}
{\cal C}_{L,i}^{H,(2)}(z,\frac{Q^2}{M^2},m^2) = 16\pi\frac{Q^2}{m^2z}
\{c_{L,i}^{(1)}(\eta,\xi) + \bar c_{L,i}^{(1)}(\eta,\xi)\ln\frac{M^2}{m^2}
\} \,,
\end{eqnarray}
\begin{eqnarray}
{\cal C}_{2,i}^{H,(2)}(z,\frac{Q^2}{M^2},m^2) &=& 16\pi\frac{Q^2}{m^2z}
\{c_{T,i}^{(1)}(\eta,\xi) + c_{L,i}^{(1)}(\eta,\xi)
+[\bar c_{T,i}^{(1)}(\eta,\xi) \nonumber \\ && \qquad 
+ \bar c_{L,i}^{(1)}(\eta,\xi)
]\ln\frac{M^2}{m^2} \}\,.
\end{eqnarray}
In the above expressions the coefficients
$c_{k,i}^{(1)}$, $\bar c_{k,i}^{(1)}$ 
and $d_{k,i}^{(1)}$ for $k=T,L$ and $i=q,g$ are defined in eqns.(5.3)-
(5.6) of \cite{lrsn1}. As has already been mentioned they are too
long to be presented in a paper and they are available upon request.
Like the coefficient functions in table 1 the heavy flavor
contributions can be decomposed in color factors in a similar
way. In first order in $\alpha_s$ we have
\begin{eqnarray}
{\cal C}_{k,g}^{H,(1)}(z,Q^2,m^2) = T_f
{\cal C}_{k,\gamma}^{H,(0)}(z,Q^2,m^2) \,,
\end{eqnarray}
where ${\cal C}_{k,\gamma}^{H,(0)}$ denotes the photonic coefficient which
is given in eqs.(2.27) and (2.28) (see also (2.36)).
In second order in $\alpha_s$ the expressions are analogous to the
ones presented for light quark production in (A.1),
(A,3) and (A.4)
\begin{eqnarray}
{\cal C}_{k,q}^{(2)}(z,Q^2,m^2) = T_f C_F
B_{FF}^{(k)}(z,Q^2,m^2) \,,
\end{eqnarray}
\begin{eqnarray}
{\cal C}_{k,q}^{H,(2)}(z,\frac{Q^2}{M^2},m^2) = T_f C_F
D_{FF}^{(k)}(z,\frac{Q^2}{M^2},m^2) \,,
\end{eqnarray}
and
\begin{eqnarray}
{\cal C}_{k,g}^{H,(2)}(z,\frac{Q^2}{M^2},m^2) = T_f C_F
E_{FF}^{(k)}(z,\frac{Q^2}{M^2},m^2) +
T_f C_A E_{FA}^{(k)}(z,\frac{Q^2}{M^2},m^2) 
\,.
\end{eqnarray}
Notice that in the limit $m \rightarrow 0$ the above expressions
need an additional mass factorization. After this procedure is carried
out the coefficients $B_{FF}$, $D_{FF}$, $E_{FF}$
and $E_{FA}$ pass into their massless analogues defined in (A.1), (A.3)
and (A.4). The order $\alpha_s$ contributions to the photonic coefficient
function ${\cal C}_{k,\gamma}^H$ can be derived from (A.16). It is equal to
\begin{eqnarray}
{\cal C}_{k,\gamma}^{H,(1)}(z,Q^2,m^2) =  C_F
E_{FF}^{(k)}(z,Q^2,m^2) \,,
\end{eqnarray}

\newpage
%

Figure Captions
\begin{description}
\item[Fig.1.]
The process $e^-(p_e) + e^+ \rightarrow e^-(p_e') + e^+ + X$,
where $X$ denotes any hadronic state.
\item[Fig.2.]
The lowest order Feynman diagrams contributing to the
Born reaction $\gamma^*(q) + \gamma(k) \rightarrow q + \bar q$.
\item[Fig.3.]
Feynman diagrams contributing to the one-loop correction to the
process $\gamma^*(q) + \gamma(k) \rightarrow q + \bar q$.
Additional graphs are obtained by reversing the arrows on
the quark lines. Graphs containing the external quark self-energies
are included in the calculation but not shown in the figure.
\item[Fig.4.]
The order $g$ $(\alpha_s = g^2/4\pi$) Feynman diagrams contributing
to the gluon bremsstrahlung process
$\gamma^*(q) + \gamma(k) \rightarrow q + \bar q + g$.
Additional graphs are obtained by reversing the arrows on the quark
lines.
\item[Fig.5.]
The $x$-dependence of $F^\gamma_2(x,Q^2)$ at $Q^2 = 5.9$ $({\rm GeV}/c)^2$,
solid line: $F^\gamma_2(NLO)$, 
long-dashed line: $F^\gamma_2(LO)$,
short-dashed line: NLO heavy quark contributions,
dotted line: LO heavy quark contributions.
The data are from PLUTO \cite{pluto}.
\item[Fig.6.]
The $x$-dependence of $F^\gamma_L(x,Q^2)$ at $Q^2 = 5.9$ $({\rm GeV}/c)^2$,
solid line: $F^\gamma_L(NLO)$, 
long-dashed line: $F^\gamma_L(LO)$,
short-dashed line: NLO heavy quark contributions,
dotted line: LO heavy quark contributions.
\item[Fig.7.]
The $x$-dependence at LO 
of $F_2^\gamma(x,Q^2)$ at $Q^2 = 5.9$ $({\rm GeV}/c)^2$
for three choices of the mass factorization scale $M^2$:
$M=2Q$ (long-dashed line), $M=Q$ (solid line) and 
$M=Q/2$ (short-dashed line).
The data are from PLUTO \cite{pluto}.
\item[Fig.8.]
The $x$-dependence at LO 
of $F_L^\gamma(x,Q^2)$ at $Q^2 = 5.9$ $({\rm GeV}/c)^2$
for three choices of the mass factorization scale $M^2$:
$M=2Q$ (long-dashed line), $M=Q$ (solid line) and 
$M=Q/2$ (short-dashed line).
\item[Fig.9.]
The $x$-dependence at NLO
of $F_2^\gamma(x,Q^2)$ at $Q^2 = 5.9$ $({\rm GeV}/c)^2$
for three choices of the mass factorization scale $M^2$:
$M=2Q$ (long-dashed line), $M=Q$ (solid line) and
$M=Q/2$ (short-dashed line).
The data are from PLUTO \cite{pluto}.
\item[Fig.10.]
The $x$-dependence at NLO
of $F_L^\gamma(x,Q^2)$ at $Q^2 = 5.9$ $({\rm GeV}/c)^2$
for three choices of the mass factorization scale $M^2$:
$M=2Q$ (long-dashed line), $M=Q$ (solid line) and
$M=Q/2$ (short-dashed line).
\item[Fig.11]
The $x$-dependence of the NLO massless ($n_f=3$) plus the
massive charm-quark contribution 
to $F_2^\gamma(x,Q^2)$ (solid line) compared with
the NLO massless contribution ($n_f=4$, dashed line),
at $Q^2 = 5.9$ $({\rm GeV}/c)^2$.
The data are from PLUTO \cite{pluto}.
\item[Fig.12.]
The $x$-dependence of the NLO massless ($n_f=3$) plus the
massive charm-quark contribution 
to $F_L^\gamma(x,Q^2)$ (solid line) compared with
the NLO massless contribution ($n_f=4$, dashed line),
at $Q^2 = 5.9$ $({\rm GeV}/c)^2$.
\item[Fig.13.]
The $x$-dependence of the LO and NLO massive hadronic charm 
contributions to $F_2^\gamma(x,Q^2)$ (solid lines) compared with
the LO and NLO massive photonic charm contributions (dashed lines),
at $Q^2 = 5.9$ $({\rm GeV}/c)^2$.  The NLO contributions are the larger
ones.
\item[Fig.14]
The $x$-dependence of the LO and NLO massive hadronic charm 
contributions to $F_L^\gamma(x,Q^2)$ (solid lines) compared with
the LO and NLO massive photonic charm contributions (dashed lines),
at $Q^2 = 5.9$ $({\rm GeV}/c)^2$.  The NLO contributions are the larger ones.
\item[Fig.15.]
The $x$-dependence of $F^\gamma_2(x,Q^2)$ at $Q^2 = 51$ $({\rm GeV}/c)^2$,
solid line: $F^\gamma_2(NLO)$, 
long-dashed line: $F^\gamma_2(LO)$,
short-dashed line: NLO heavy quark contributions,
dotted line: LO heavy quark contributions.
The data are from AMY \cite{amy}.
\item[Fig.16.]
The $x$-dependence of $F^\gamma_L(x,Q^2)$ at $Q^2 = 51$ $({\rm GeV}/c)^2$,
solid line: $F^\gamma_L(NLO)$, 
long-dashed line: $F^\gamma_L(LO)$,
short-dashed line: NLO heavy quark contributions,
dotted line: LO heavy quark contributions.
\item[Fig.17.]
The $x$-dependence at LO
of $F_2^\gamma(x,Q^2)$ at $Q^2 = 51$ $({\rm GeV}/c)^2$
for three choices of the mass factorization scale $M^2$:
$M=2Q$ (long-dashed line), $M=Q$ (solid line) and
$M=Q/2$ (short-dashed line).
The data are from AMY \cite{amy}.
\item[Fig.18.]
The $x$-dependence at LO
of $F_L^\gamma(x,Q^2)$ at $Q^2 = 51$ $({\rm GeV}/c)^2$
for three choices of the mass factorization scale $M^2$:
$M=2Q$ (long-dashed line), $M=Q$ (solid line) and
$M=Q/2$ (short-dashed line).
\item[Fig.19.]
The $x$-dependence at NLO
of $F_2^\gamma(x,Q^2)$ at $Q^2 = 51$ $({\rm GeV}/c)^2$
for three choices of the mass factorization scale $M^2$:
$M=2Q$ (long-dashed line), $M=Q$ (solid line) and
$M=Q/2$ (short-dashed line).
The data are from AMY \cite{amy}.
\item[Fig.20.]
The $x$-dependence at NLO
of $F_L^\gamma(x,Q^2)$ at $Q^2 = 51$ $({\rm GeV}/c)^2$
for three choices of the mass factorization scale $M^2$:
$M=2Q$ (long-dashed line), $M=Q$ (solid line) and
$M=Q/2$ (short-dashed line).
\item[Fig.21]
The $x$-dependence of the NLO massless ($n_f=3$) plus the
massive charm-quark contribution 
to $F_2^\gamma(x,Q^2)$ (solid line) compared with
the NLO massless contribution ($n_f=4$, dashed line),
at $Q^2 = 51$ $({\rm GeV}/c)^2$.
The data are from AMY \cite{amy}.
\item[Fig.22.]
The $x$-dependence of the NLO massless ($n_f=3$) plus the
massive charm-quark contribution 
to $F_L^\gamma(x,Q^2)$ (solid line) compared with
the NLO massless contribution ($n_f=4$, dashed line),
at $Q^2 = 51$ $({\rm GeV}/c)^2$.
\item[Fig.23.]
The $x$-dependence of the LO and NLO massive hadronic charm 
contributions to $F_2^\gamma(x,Q^2)$ (solid lines) compared with
the LO and NLO massive photonic charm contributions (dashed lines),
at $Q^2 = 51$ $({\rm GeV}/c)^2$.  The NLO contributions are the larger ones.
\item[Fig.24]
The $x$-dependence of the LO and NLO massive hadronic charm 
contributions to $F_L^\gamma(x,Q^2)$ (solid line) compared with
the LO and NLO massive photonic charm contributions (dashed lines),
at $Q^2 = 51$ $({\rm GeV}/c)^2$.  The NLO contributions are the larger
ones.
\end{description}


\begin{thebibliography}{99}
\bibitem{pluto} Ch. Berger {\it et al}., (PLUTO Collaboration), 
Phys. Lett. \underline{B107}, 168 (1981), 
ibid \underline{B142}, 111 (1984),
ibid \underline{B149}, 421 (1984),
Z. Phys \underline{C26}, 353 (1984),
Nucl. Phys. \underline{B281}, 365 (1987).
\bibitem{cello}
H.J. Behrend {\it et al}.,
(CELLO Collaboration), Phys. Lett. \underline{B126}, 391 (1983);
C. Kiesling, contributed paper to the {\it XXV International Conference
on High Energy Physics}, Editors  K.K. Phua and Y. Yamaguchi,
South East Asia Theoretical Physics Association
and the Physical Society of Japan, Singapore, (1990), unpublished.
\bibitem{tpc}
H. Aihara {\it et al}., 
(TPC2$\gamma$ Collaboration) Phys. Rev. Lett. \underline{58}, 97 (1987);
Z. Phys \underline{C34}, 1 (1987);
D. Bintinger {\it et al}., Phys. Rev. Lett. \underline{54}, 763 (1985).
\bibitem{tasso}
M. Althoff {\it et al}., (TASSO Collaboration), Z. Phys. 
\underline{C31}, 527 (1986);
\bibitem {jade} W. Bartel {\it et al}., (JADE Collaboration)
Z. Phys. \underline{C24}, 231 (1984); Phys. Lett. \underline{B121}, 205
(1983).
\bibitem{amy} R. Tanaka (AMY Collaboration) in 
{\it IX International Workshop on Photon-Photon
Collisions}, March 22-26, 1992, University of California, San Diego,
California. KEK preprint 92-37; 
R. Tanaka {\it et al}., Phys.Lett. \underline{B277}, 215 (1992);
T. Sasaki {\it et al}., Phys.Lett. \underline{B252}, 491 (1990);
T. Nozaki in {\it  Proceedings of the 
Joint International Lepton-Photon Symposium and Europhysics
Conference on High Energy Physics}, eds S. Hegarty, K. Potter,
and E. Quercigh,  World Scientific, (1991) p.156.
\bibitem{venus}M. Chiba (VENUS Collaboration) in 
{\it Proceedings of the XXVIth
Rencontres de Moriond (High Energy Hadronic Interactions)}
Les Arcs, March, 1991.
\bibitem{topaz} H. Hayashii (TOPAZ collaboration) in 
{\it IX International Workshop on Photon-Photon
Collisions}, March 22-26, 1992, University of California, San Diego,
California.
\bibitem{ali} A. Ali {\it et al}., {\it Physics at LEP} 
edited by J. Ellis and 
R. Peccei, CERN-86-02 Geneva (1986) vol.2, p 81.
\bibitem{dg1}M. Drees and R.M. Godbole, DESY 92-044; Phys. Rev. Lett.
\underline{67}, 1189 (1991).
\bibitem{eghn} O.J.P. \'Eboli, M.C. Gonzalez-Garcia, F. Halzen
and S.F. Novaes, Phys. Rev. \underline{D47}, 1889 (1993);
M. Drees, M. Kr\"amer, J. Zunft and P.M. Zerwas, DESY 92-169.
\bibitem{ew} E. Witten, Nucl. Phys. \underline{B120},189 (1977).
\bibitem{bill}W.A. Bardeen in {\it Proceedings of the 1981
International Symposium on Lepton and Photon Interactions
at High Energies, Bonn}, edited by W. Pfeil (Physikalisches Institut,
Universit\"at Bonn, Bonn 1981), p 432.	
\bibitem{gr}M. Gl\"uck and E. Reya, Phys. Rev. \underline{D28},
2749 (1983).
\bibitem{jf} J.H. Field in {\it VIII
International Workshop on Photon-Photon Collisions, Jerusalem,
April (1988)},
World Scientific, (1989) p.349; S. Kawabata, in {\it Proceedings of the 
Joint International Lepton-Photon Symposium and Europhysics
Conference on High Energy Physics}, eds S. Hegarty, K. Potter,
and E. Quercigh,  World Scientific, (1991) p.53.
\bibitem{dg}M. Drees and K. Grassie, Z. Phys. \underline{C28},451 (1985).
\bibitem{acl}H. Abramowicz, K. Charchula and A. Levy, Phys. Lett.
\underline{B269}, 458 (1991);
see also H. Abramowicz et al., DESY 91-057.
\bibitem{grv1} M. Gl\"uck, E. Reya and A. Vogt, Phys. Rev.
\underline{D45}, 3986 (1992).
\bibitem{grv2} M. Gl\"uck, E. Reya and A. Vogt, Phys. Rev.
\underline{D46}, 1973 (1992).
\bibitem{acfgp}P. Aurenche, P. Chiapetta, M. Fontannaz, J.Ph. Guillet
and E. Pilon, preprint LPTHE-Orsay 92/13.
\bibitem{gs} L.E. Gordon and J.K. Storrow, Z. Phys. \underline{C56},
307 (1992).
\bibitem{fkp} J.H. Field, F. Kapusta and L. Poggioli, 
Z. Phys. \underline{C36}, 121 (1987);Phys. Lett. \underline{181},
362 (1986); J.H. Da Luz Vieira and J.K. Storrow, Z. Phys. 
\underline{C51}, 241 (1991).
\bibitem {dgr} M. Drees, M. Gl\"uck and E. Reya, Phys. Rev. 
\underline{D30}, 2316 (1984).
\bibitem{zn} E.B. Zijlstra and W.L. van Neerven, 
Nucl. Phys. \underline{B383},525 (1992);
Phys. Lett. \underline{B273},476 (1991);
W.L. van Neerven and E.B. Zijlstra, Phys.Lett.
\underline{B272}, 127 (1991).
\bibitem{lrsn1} E. Laenen, S. Riemersma, J. Smith and W.L. van Neerven,
Nucl. Phys. \underline{B392}, 162 (1993); 
ibid \underline{B392}, 229 (1993);
Phys. Lett. \underline{B291}, 325 (1992).
\bibitem{hr} C.T. Hill and G.G Ross, Nucl. Phys. \underline{B148},
373 (1979).
\bibitem{sn} J. Smith and W.L. van Neerven,
Nucl.Phys. \underline{B374},36 (1992).
\bibitem {ggr1} M. Gl\"uck, K. Grassie and E. Reya, Phys. Rev. 
\underline{D30}, 1447 (1984).
\bibitem{bb}W.A. Bardeen and A.J. Buras, Phys. Rev. \underline{D20},
166 (1979); E \underline{D21} 2041 (1980). 
\bibitem{fp}M. Fontannaz and E. Pilon, Phys. Rev. \underline{D45},
382 (1992).
\bibitem{ew1} E. Witten, Nucl.Phys. \underline{B104},445 (1976);
J. Babcock and D. Sivers, Phys.Rev. \underline{D18},2301 (1978);
M.A. Shifman, A.I. Vainstein and V.J. Zakharov,
Nucl.Phys.\underline{B136}, 157 (1978);
M. Gl\"uck and E. Reya, Phys.Lett. \underline{83B}, 98 (1979);
J.V. Leveille and T. Weiler, Nucl.Phys. \underline{B147}, 147 (1979).
\bibitem{mg} M. Gl\"uck in {\it Proceedings of the HERA Workshop},
Hamburg. Oct. 1987 ed. R.D. Peccei, vol. 1, p.119;
M. Gl\"uck, R.M. Godbole and E.Reya, Z.Phys. \underline{C38}, 441 (1988).
\end{thebibliography}
\end{document}